\documentclass[modern]{aastex61}

\usepackage{graphicx}
\usepackage{amsmath}
\usepackage{amssymb}
\usepackage{enumitem}

\newcommand\mybar{\kern1pt\rule[-\dp\strutbox]{.8pt}{\baselineskip}\kern1pt}

\setlist[itemize]{noitemsep, topsep=0pt, leftmargin=*}

\shorttitle{Extinction of the TeV Background by Sunlight}
\shortauthors{Loeb}

\begin{document}

\title{Extinction of the TeV Gamma-Ray Background by Sunlight}

\author{Abraham Loeb}
\affiliation{Astronomy Department, Harvard University, 60 Garden
  St., Cambridge, MA 02138, USA}

\begin{abstract}
I show that pair production on sunlight introduces a sizable
anisotropy in the cosmic background of TeV gamma-rays. The anisotropy
amplitude in the direction of the Sun exceeds the cosmic dipole
anisotropy from the motion of the Sun relative to the cosmic
rest-frame.
\end{abstract}

\section{Introduction}

Given its effective surface temperature of $T_{\rm eff}=5.8\times
10^3~{\rm K}$, the Sun acts as a source of photons with a
characteristic energy, $E_{{\gamma}_\odot}\approx k_BT_{\rm eff}=
0.5~{\rm eV}$. When these photons collide with the cosmic background
of TeV photons \citep{2016ApJ...818..187I,2022ChA&A..46...42Y}, they
produce electron-positron pairs through the process: $\gamma_\odot
+ \gamma\rightarrow e^+ + e^-$. For pair creation to occur, the energy
of the background photon, $E_\gamma$, must exceed the threshold value,
$E_{\rm th}$ \citep{1967PhRv..155.1404G},
\begin{equation}
E_{\gamma}>E_{\rm th}=\frac{2m_e^2c^4}{E_{{\gamma}_\odot}\mu}\approx
{0.5~{\rm TeV}\over (\mu/2)},
\label{one}
\end{equation}
where $m_e=5.11\times 10^5~{\rm eV}$ is the electron rest-mass, $c$ is
the speed of light and $\mu=(1-\cos\theta)$ with $\theta$ being the
angle between the momenta of the two photons in the observer's frame.
The value of $E_{\rm th}$ is minimized for a head-on collision
($\mu=2$) with a background photon of energy $E_\gamma\gtrsim 0.5~{\rm TeV}$.

Above this threshold, the cross-section for pair production is given by,
\begin{equation}
\sigma_{\gamma\gamma}= 3.7\times 10^{-25}~{\rm cm^2}\times
(1-\beta^2)\left[(1-{1\over 3}\beta^4)\ln\left({{1+\beta}\over
    {1-\beta}}\right)-{2\over 3}\beta(2-\beta^2)\right],
\label{two}
\end{equation}
where $\beta=(v/c)$, with $v$ being the speed of the electron or
positron in the center-of-mass system. The value of $\beta$ depends on
$E_\gamma$ and $\theta$ through the relation,
\begin{equation}
\beta=\sqrt{1- {E_{\rm th}\over E_\gamma}}.
\label{three}
\end{equation}

The same pair-production process introduces a cosmic opacity to the
TeV radiation background as a result of photon-photon collisions with
the extragalactic background light
\citep{1967PhRv..155.1408G,2021Ap&SS.366...51S,2021Univ....7..146F}. Here
we focus on the imprint of Solar photons on the cosmic TeV background
observable on Earth.

\section{Results}

The number density of Solar photons at a distance $R$ from the Sun is
given by,
\begin{equation}
n_{{\gamma}_\odot}= {L_\odot\over \left(4\pi c R^2
  E_{{\gamma}_\odot}\right)}\approx 2.7\times 10^{12}~{\rm
  cm^{-3}}\left({R\over R_\odot}\right)^{-2},
\label{four}
\end{equation}
where $L_\odot = 4\times 10^{33}~{\rm erg~s^{-1}}$ is the Solar
luminosity. The Earth-Sun separation at $R=1.5\times 10^{13}~{\rm cm}$
is $2.14\times 10^2$ times larger than the Solar radius
$R_\odot=7\times 10^{10}~{\rm cm}$. This photon density yields a local
optical-depth for pair-production at an energy $E_{\gamma}\sim 1~{\rm
  TeV}$, of magnitude,
\begin{equation}
\tau_{\gamma\gamma}(E_\gamma\sim 1~{\rm TeV})=
n_{{\gamma}_\odot}\sigma_{\gamma\gamma}R \sim 7\times
10^{-2}\left({R\over R_\odot}\right)^{-1} .
\label{four}
\end{equation}

The resulting absorption of background photons produces a deficit in
the gamma-ray background at the energy $\sim 1~{\rm TeV}$ with an
anisotropy amplitude $\sim 7\times 10^{-2}$ in the vicinity of the
Sun and $\sim 3.3\times 10^{-4}$ at the Earth's orbital radius, and an
anisotropy angular dependence that reflects the position of the Sun in
the sky at the observing time combined with the $\mu$ dependence of
$\sigma_{\gamma\gamma}$ in equations~(\ref{two}-\ref{three}).

\section{Implications}

Interestingly, the expected anisotropy amplitude is significant
relative to the dipole associated with the motion of the Solar system
relative to the cosmic rest-frame, $\sim 1.2\times 10^{-3}$, as
measured from the cosmic microwave background
\citep{1993ApJ...419....1K}.

\bigskip
\bigskip
\section*{Acknowledgements}

This work was supported in part by Harvard's {\it Black Hole
  Initiative}, which is funded by grants from JFT and GBMF.  

\bigskip
\bigskip
\bigskip

\bibliographystyle{aasjournal}
\bibliography{t}
\label{lastpage}
\end{document}